\documentclass[pre,aps,twocolumn]{revtex4}

\usepackage{epsfig}
\usepackage{bm}
\newcommand{\B}[1]{{\bm{#1}}} \newcommand{\C}[1]{{\mathcal{#1}}}
\newcommand{\Onecol} {\begin{widetext} \onecolumngrid} 
\newcommand{\Twocol} {\end{widetext} \twocolumngrid} 
 \newcommand{\be}{\begin{equation}}
\newcommand{\ba}{\begin{array}} \newcommand{\bea}{\begin{eqnarray}}
\newcommand{\bfi}{\begin{figure}} \newcommand{\ee}{\end{equation}}
\newcommand{\ea}{\end{array}} \newcommand{\eea}{\end{eqnarray}}
 \newcommand{\efi}{\end{figure}}

 \def\Fbox#1{\vskip1ex\hbox to 8.5cm{\hfil\fboxsep0.3cm\fbox{%
  \parbox{8.0cm}{#1}}\hfil}\vskip1ex\noindent}  
 \begin{document} \title{Strong Universality in Forced and Decaying
 Turbulence}
  \author{Victor S. L'vov$^1$, Ruben Pasmanter$^
{1,2}$, Anna Pomyalov$^1$ and
  Itamar Procaccia$^{1,3}$}
  \affiliation{$^1$Department of Chemical Physics,
  The Weizmann Institute of
  Science, Rehovot 76100, Israel.\\
$^2$  Royal Dutch Meteorological Institute,
 POB 201, 3730 AE De Bilt, Netherlands\\
$^3$ Department of Physics, the Chinese University of Hong Kong,
 Shatin, Hong Kong.
} \pacs{}
\begin{abstract}
The weak version of universality in turbulence refers to the
independence of the scaling exponents of the $n$th order strcuture
functions from the statistics of the forcing. The strong version
includes universality of the coefficients of the structure
functions in the isotropic sector, once normalized by the mean
energy flux. We demonstrate that shell models of turbulence
exhibit strong universality for both forced and decaying
turbulence. The exponents {\em and} the normalized coefficients
are time independent in decaying turbulence, forcing independent
in forced turbulence, and equal for decaying and forced
turbulence. We conjecture that this is also the case for
Navier-Stokes turbulence.
\end{abstract}
\maketitle

\section{Introduction}

The statistical theory of fluid turbulence is concerned with
correlation functions of the turbulent velocity vector field $\B
u(\B r,t)$ where $\B r$ is the spatial position and $t$ the time
\cite{75MY}. Since the velocity field is a vector, multi-point and
multi-time correlation functions are in general tensor functions
of the vector positions and the scalar times. Naturally such
functions have rather complicated forms which are difficult to
measure and to compute. Consequently, almost from its very
beginning, the statistical theory of turbulence had been
discussed in the context of an isotropic and homogeneous model.
The notion of isotropic turbulence was first introduced by
G.~I.~Taylor in 1935 \cite{35Tay}. It refers to a turbulent flow,
in which the statistical averages of every function of the
velocity field and its derivatives with respect to a particular
frame of axes is invariant to any rotation in the axes. This is a
very effective mathematical simplification which, if properly
used, can drastically reduce the mathematical complexity of the
theory. For this reason, it was very soon adopted by others, such
as T.~D.~K\'arm\'an and L.~Howarth \cite{38KH} who derived the
K\'arm\'an-Howarth equation, and A.~N.~Kolmogorov \cite{41Ka,
41Kb} who derived the 4/5 law. In fact, most of the theoretical
work in turbulence in the past sixty years had been limited to
the isotropic model.

Within the homogeneous and isotropic model there developed the
notion of universality of turbulence. By universality, we mean the
tendency of different turbulent systems to show, for very large
Reynolds numbers $ \C R e$, the same small-scales statistical
behavior when the measurements are done far away from the
boundaries. The statistical objects (see bellow for definitions)
exhibit approximately the same scaling exponents whether they are
measured in the atmospheric boundary layer, in a wind tunnel or
in a computer simulation, provided they are measured far from the
boundaries. Moreover, the accumulated experimental knowledge over
the years indicated that not only in forced, stationary
turbulence, but also in decaying turbulence, there is a regime of
time where the statistical objects exhibit the same scaling
properties. This phenomenon was explained \cite{75MY} by the
widely separated time scales (``eddy turn over times") that
characterize large and small length scales in turbulence. While
turbulence was decaying on the time scale of the large eddies,
the small one had ample time to reach an ``energy-flux
equilibrium" that in terms of scaling behavior was
indistinguishable from forced turbulence. Thus there exists a
wide-spread belief that at least from the point of view of
scaling exponents, forced and decaying turbulence are in the same
universality class, sharing the same scaling exponents of the
corresponding correlation functions.

To actually {\em prove} this type of universality in experiments
and simulations is however far from straightforward. To achieve
reasonable precision in the measurement of scaling exponents one
needs large ranges of scales where scaling prevails, and this
entails large Reynolds numbers. Unfortunately large Reynolds
numbers are available usually when anisotropic effects are large,
like in the atmospheric boundary layer or in large wind tunnels.
Direct Numerical Simulations (DNS) can be used to eliminate
anisotropy almost completely (up to lattice anisotropy which are
unavoidable in simulations), but they are limited to relatively
low $\C Re$, notwithstanding the very recent simulations of size
$4096^3$ \cite{02Kan}. Decaying turbulence is even harder to
characterize precisely, since the effective Reynolds number
decreases in time. Thus actual measurements of scaling properties
are fraught with difficulties, corrections to scaling, effects of
anisotropy and what not. As a result, over the years \cite{95BO}
and also very recently, it was proposed \cite{00FONGY} that
structure functions in forced and decaying turbulence have
different exponents. In this paper we take a strong stand,
proposing that the universality that actually exists in
turbulence is even stronger than what has been anticipated so far.

To make our point clear, recall that the statistical description
of fully developed turbulence employs correlation functions and
structure functions. These are ensemble average of velocity
differences across a length-scale $R$. In the theoretical studies
of turbulence the two most common ensemble averages are {\em over
realizations of the forcing} when one studies forced turbulence,
or {\em over initial conditions} when one studies decaying
turbulence. The longitudinal structure functions are the simplest
such objects, being moments of the longitudinal components of the
velocity difference between two points. We will denote the
longitudinal structure functions in forced and decaying turbulence
by $S_p$ and $F_p$ respectively, with the precise definitions
\begin{eqnarray} S_p(R)\equiv \langle \{ [\B u(\B r+\B R,t)- \B
u(\B r,t)]\cdot \frac{\B R}{R} \} ^p\rangle_{\rm f}\ , \label{Sn}\\
F_p(R,t)\equiv \langle \{ [\B u(\B r+\B R,t)- \B u(\B r,t)]\cdot
\frac{\B R}{R} \} ^p\rangle_{\rm i}\ .  \label{Fn}
\end{eqnarray}
Here $\B u(\B r,t)$ is the velocity field measured at point $\B r$
at time $t$. $\langle \dots \rangle_{\rm f}$ and $\langle \dots
\rangle_{\rm i}$ stand for ensemble averaging over the forcing and
the initial conditions respectively. In writing Eq.~(\ref{Sn}) we
assumed that the forcing is stationary in time, homogeneous and
isotropic, and thus $S_n$ is a function of the scalar $R$ only. In
writing Eq.~(\ref{Fn}) we assumed that the initial condition are
homogeneous and isotropic. Of course the decaying structure
functions are by definition time dependent. The widely spread
belief \cite{75MY,53Bat,95Fri} is  for $R$ values in the inertial
range of turbulence (much smaller than the forcing scale but much
larger than the dissipation scale), the scaling exponents
$\zeta_p$  that characterize $S_p(R)$, i.e. $S_p(\lambda R)
=\lambda^{\zeta_p} S_p(R)$, are the same as the scaling exponents
that characterize $F_p(R,t)$ for a given value of $t$. Of course
also here $R$ should be well in the (time dependent) inertial
range and $t$ should be neither too small nor too large
\cite{00SS}. In the sequel we refer to the identity of the only
scaling exponents of these two sets of objects (if it exists) as
``the weak version of universality". As mentioned above the
existence of the weak version of universality is by no means
accepted by everybody in the field of turbulence. Since there is
no {\em proof} of this universality, doubts of its existence
linger, and for example in \cite{00FONGY} it was concluded that
the scaling exponents of the two families of statistical objects
are {\em not} the same. We note however that in the same paper it
was stated that the scaling exponents of the longitudinal and
transverse structure functions are also not the same. It was shown
recently however that such statements stem from incomplete
treatments of the effects of anisotropy \cite{98ADKLPS,00KLPS},
leaving hope that the weak version of universality is still
correct.

In fact, in this paper we will propose that not only the weak
version of universality is correct, but in fact also a ``strong
version of universality" is applicable. By the latter we mean that
once properly normalized, the structure functions $F_p$ and $S_p$
agree not only in exponents but also in amplitudes. In the context
of the 2nd order structure function this is not a new statement.
The universality of ($\zeta_2$ and $C_2$) was already stated in
the 80's by Yaglom~\cite{81Yag} and Kader~\cite{84Kad}. Analyzing
hundreds of experiments made in different flows under different
conditions, Sreenivasan in 1995 came  to the conclusion that ``the
Kolmogorov constant  $C_2$  is \emph{more or less} universal,
essentially independent of the flow as well as the Reynolds number
(for $R_\lambda > 50$ or so),.. with the average value of
$C_2\approx 0.53$ with a standard deviation of about
0.055"~\cite{95Sre}. Nevertheless the universality of $C_2$ and
$C_p$ for $p\le 4$ is still under debate. For example, very
recently in~\cite{03BBCLTV} the authors argued on the basis of a
$256^3$ DNS ``in favor of an ``exponents only" universality
scenario for forced turbulence".

We believe that this strong version of universality was never
stated before, and the common thinking is that amplitudes depend
in a non-universal way on details of the forcing or the
preparation of the decaying turbulence. While true, we will argue
that the freedom afforded by such details is very limited,
amounting at the end to just {\em one} free number, which, once
taken into account, the strong version of universality applies.

Besides being an issue of fundamental importance to turbulence,
there is another reason for returning at this time to the
correspondence between force and decaying turbulence. The reason
is that the riddle of anomalous scaling of correlation and
structure functions in forced turbulent advection (passive and
active) had been solved recently. First in the context of the
non-generic Kraichnan model of passive scalar advection
\cite{01FGV}, and then, in steps, for passive vectors
\cite{96Ver,00AP}, generic passive scalars and vectors
\cite{01CV,01ABCPV,02CGP} and finally for generic active scalar
and vectors \cite{01CMMV,02CCGP,03CCGP}. The common thread of this
advance is that anomalous scaling is discussed in the context of
the decaying (unforced problem), in which one shows that there
exist Statistically Preserved Structures (eigenfunctions of
eigenvalue 1 of the appropriate propagator of the decaying
correlation functions). The decaying problem is independent of
forcing, and one shows that the statistics of the forced problem
is dominated by the same Statistically Preserved Structures that
are identified in the decaying problem. The calculation of the
anomalous exponents boils down then to calculating eigenfunctions
of linear operators. In these problems the correspondence between
the decaying and forced statistics is proven mathematically or
demonstrated beyond reasonable doubt by careful numerics. A
crucial ingredient in all this progress is that turbulent
advection is described by {\em linear} pde's. There is therefore
an urgent question how to translate (if it is possible) the newly
acquired insights to the non-linear turbulent problem itself, be
it the Navier-Stokes equations or any of the shell models that
were frequently discussed recently in the context of anomalous
scaling. In this paper we make a step in this direction, analyzing
the decay of the Sabra  shell model \cite{98LPPPV} and showing
numerically that the statistics of the decaying state and the
forced turbulent state {\em are the same} in exponents {\em and in
amplitudes} up to one freedom (the time dependent mean energy). We
opt to work with the shell model rather than the Navier-Stokes
equations simply because the accuracy required for our aims
exceeds the available scaling ranges and decay times for the
latter. We express a strong belief that very similar results can
be demonstrated also for Navier-Stokes turbulence. Indeed, in a
future publication we will present the theory that stands behind
the present numerical findings and demonstrate that the basic
structure of that theory is the same for shell models and the
Navier-Stokes equations.

The paper is organized as follows: in Sec.~\ref{s:model} we
introduce the shell model and the numerical
 simulations that we perform. We present the data for the energy
 decay and explain what is the time domain for which we should
  compare the decaying and the forced statistics. In Sec.~\ref{s:forced}
 we present the results for forced structure functions for
  different types of forcing. In determining the exponents {\em
  and the amplitudes} of these functions one has to be extra
  careful - we explain that one needs to find fits to functions
  throughout their range of existence. It is not enough to plot
  log-log plots for the inertial range. In Sect.~\ref{s:decay}
   we present the data for the decaying correlation functions, and
 explain how to find their exponents and amplitudes once the time
 dependence is taken into account. We explain theoretically that
 the decaying structure functions contain sub-leading
 contributions that decay fast toward small scales and do not
 affect the leading  scaling exponents.
 The results of our calculations are summarized in Table 1 which
 is the central result of this paper, giving strong support to
  the conjecture of {\em strong} universality.
  In Sec.~\ref{s:sum} we present a summary and some
  concluding remarks.

\section{Model and energy decay}
\label{s:model}
\subsection{Model and objectives}
The Sabra  shell model \cite{98LPPPV}, like all shell models, is a
reduced dynamical description of turbulence in terms of complex
variables $u_n$ which represent velocity amplitudes associated
with wavenumber $k_n=k_0\lambda^n$.  The equations of motion are
\begin{eqnarray} \label{98LPPPV} \frac{d u_n}{dt}&=&i\big( ak_{n+1}
u_{n+2}u_{n+1}^*
 + bk_n u_{n+1}u_{n-1}^*  \\ \nonumber &&
-ck_{n-1} u_{n-1}u_{n-2}\big)  -\nu k_n^2  u_n +f_n\,,
\end{eqnarray}
where the symbol $^*$ stands for complex conjugation and $\nu$ is
the ``viscosity". The coefficients $a,b$ and $c$ are chosen such
that $a+b+c=0$. This  guarantees the conservation of the ``energy"
\begin{equation}\label{energy} E=\sum_n |u_n|^2\,,
\end{equation} in the inviscid ($\nu=0$) forceless
limit. As it is well known, the Sabra model has a second
quadratic invariant, analogous to the helicity in fluid
mechanics, of the form
\begin{equation}
H= \sum_n (a/c)^n |u_n|^2 \ .
\end{equation}

In this paper we will compare the statistics of the forced
solution (with the forcing $f_n$ restricted to the first and
second shells, $n=1,2$) to the statistics of the decaying problem
with $f_n=0$ for all $n$. The comparison will be
 presented in
terms of the time-independent forced structure functions $S_n$ and
time dependent decaying structure functions $F_n$ defined as
follows: \begin{eqnarray}\nonumber   S_2(k_n) &\equiv& \langle
|u_n|^2\rangle_{\rm f}\ ; \qquad   F_2(k_n,t)\equiv \langle
|u_n|^2\rangle_{\rm i}\ , \\ \label{defs} S_3(k_n) &\equiv& {\rm
Im}\langle u_{n-1}u_nu^*_{n+1}\rangle_{\rm
f}\ ;\\
\ F_3(k_n,t)&\equiv& {\rm Im}\langle u_{n-1}u_nu^*_{n+1}\rangle_{\rm i}\ ,
\nonumber\\
S_4(k_n) &\equiv&\langle |u_n|^4\rangle_{\rm f}\ ; \qquad
F_4(k_n,t)\equiv \langle |u_n|^4\rangle_{\rm i}\ ,
\nonumber\\S_6(k_n) &\equiv& \langle |u_n|^6\rangle_{\rm f}\ ;
\qquad F_6(k_n,t)\equiv \langle |u_n|^6\rangle_{\rm i}\ .\nonumber
\end{eqnarray}
Here $\langle\dots\rangle_{\rm f}$ and $\langle \dots\rangle_{\rm
i}$ represent averaging with respects to realizations of the
forcing and the initial condition respectively for the forced and
decaying problem.

The main result of the present work is that in this model there
exists \emph{strong universality}. This  means that in the bulk of
the inertial interval  the ``decaying" structure functions
$F_p(k_n,t)$ take on the form
\begin{equation}\label{eq:scaling-Fp}
F_p(k_n,t)=C_p \left[\frac{\bar \varepsilon_{\rm i}
(t)}{k_0}\right]^{p/3}\lambda^{- n\,\zeta_p}\,,
\end{equation}
with \emph{the same anomalous scaling exponents} $\zeta_p$ and
\emph{the same dimensionless constants } $C_p$, as in the scaling
laws of the ``forced" structure functions
\begin{equation}\label{eq:scaling-Sp}
S_p(k_n,t)=C_p \left[\frac{\bar \varepsilon_{\rm
f}}{k_0}\right]^{p/3}\lambda^{- n\,\zeta_p}\ .
\end{equation}
Here
\begin{equation}\label{eq:bar-ef}
 \bar \varepsilon_{\rm i} (t) \equiv
\langle \varepsilon_n (t) \rangle_{\rm i}\,, \quad \bar
\varepsilon_{\rm f} \equiv \langle \varepsilon_n (t) \rangle_{\rm
f}\ .
\end{equation}
In these equations the instantaneous value of the energy
 flux going through $n$th
shell (in a given realization) is
\begin{eqnarray}\label{eq:flux}
\varepsilon_n (t) = 2 k_n \Big[&-&a \,\lambda \,{\rm Im}\{
u_{n}(t)u_{n+1}(t)u_{n+2}^*\}\\ \nonumber &+&c \,{\rm Im}\{
u_{n-1}(t)u_{n}(t)u_{n+1}^*\} \Big] \ .
\end{eqnarray}
(for more details, see \cite{98LPPPV}). The only difference
between Eqs. (\ref{eq:scaling-Fp}) and (\ref{eq:scaling-Sp}) is
that the energy flux $ \bar \varepsilon_{\rm i}(t)$, averaged over
the statistics of the initial conditions, is decaying in time,
while the energy flux $ \bar \varepsilon_{\rm f}$, averaged over
the statistics of the forcing, is time independent. Eqs.
(\ref{eq:scaling-Fp}) and (\ref{eq:scaling-Sp}) imply that the
probability distribution function (PDF) of the velocity
fluctuations $u_n$ on scales $k_n$ within the inertial range in
decaying turbulence can be obtained from the corresponding PDF in
stationary forced turbulence (and vise versa). This is achieved
simply by the interchange $ \bar \varepsilon_{\rm i}(t)
\Leftrightarrow \bar \varepsilon_{\rm f}$. Moreover, strong
universality means that the PDF dependence on $\bar
\varepsilon_{\rm i}(t)$  for the fine scales in decaying
turbulence is independent of the initial conditions, providing
that the Reynolds number $\C R{\rm e}\gg 1$ and all the initial
energy is concentrated in the region of large scales (first
shells). Similarly, the PDF dependence on $\bar \varepsilon_{\rm
f}$ for the fine scales in forced turbulence is independent of the
statistics of forcing for $\C R{\rm e}\gg 1$, provided that the
forcing is concentrated in the region of large scales.

The rest of this paper is devoted to substantiating these (strong)
propositions, which can be summarized  in the terms of  the
existence of a probability distribution function
\begin{equation}\label{rescale}%
\C P(u_{1},\ldots,u_{N},t)=\left[ v(t)\right]^{-N}P(x_{1}
,\ldots,x_{N})\,,
\end{equation}
in which  $x_{i}\equiv u_{i}/v(t)$ and $v(t)\propto [\epsilon_n
(t)]^{1/3}$ is the corresponding velocity scale.
\subsection{Simulations}
The calculations presented bellow were carried out for the Sabra
model with 28 shells, $\lambda=2,$ $a=1,$ $b=c=-0.5,$ $k_0=2^{-4}$
and $\nu=10^{-7}$.  In our simulations we employed two different
types of forcing. We denote them as Forced 1 and Forced 2. Forced
1 has white noise added to the equation of the first shell. Forced
2 is forced by a Gaussian force on the first shell which is
correlated exponentially in time. In both cases the amplitude of
the force in the first shell was chosen $f_1=0.01$, while the
forcing amplitude in the second shell was adjusted to reduce the
helicity input $f_2=\sqrt{(-c/a)}f_1$, for more details, see
\cite{98LPPPV}, pp. 1813 and 1815.

In the decaying case, the  total initial energy  $E_0$ in the two
first shells was kept constant. The amplitude of the first two
shell velocities were defined as $|u_1(0)|^2=\alpha E_0$ and
$|u_2(0)|^2=(1-\alpha)E_0$ with $\alpha$ random, uniformly
distributed in the interval $[0,1]$. The phases in both shells
were random, uniformly distributed in the interval $[0,2\pi]$.
 A $4^{ \rm th}$ order Runge-Kutta scheme with adaptive time step was
applied. The total energy decay was followed for 9 decades in
time. The statistical objects were accumulated during 5 decades in
time.

\begin{figure}
\epsfxsize=8.5cm \epsfbox{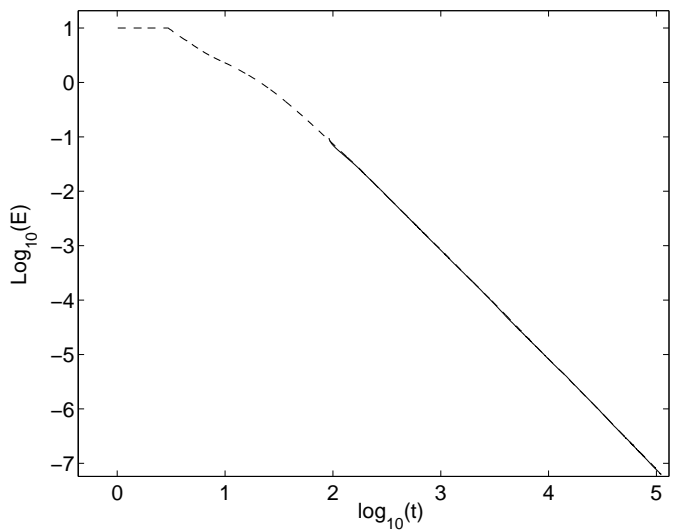}
 \caption{\label{f:2_decay} The total energy decay from Decay 1 data
(dashed line) and from Decay2 data ( solid line). The lines
coincide within the line width.}
\end{figure}

Two recording schemes for the decaying turbulence were applied. In
one case (denoted as Decay 1), the data were recorded starting
after a short transient time with $E_0=10$. For this case, the
data was averaged over 13200 initial conditions. In another case (
Decay 2), the data were recorded when the energy in each
realization have reached the value $E=0.1$ with $E_0=5$. This data
was averaged over about $37000$ initial conditions. The decay of
the total energy for two cases, plotted with an appropriate time
shift, are shown in Fig. \ref{f:2_decay}. It is clear that the two
schemes are equivalent for the study of the advanced stages of the
decay.
\subsection{The law of energy decay}
We first discuss the total energy decay, where the total energy $E(t)$
is defined as
\begin{equation}
E(t) = \sum_n |u_n|^2 \ . \label{defE}
\end{equation}
In the Navier-Stokes case the law of energy decay had been
intensively studied following the influential works of
Taylor~\cite{35Tay}, Kolmogorov~\cite{41Ka,41Kb}, Batchelor and
Townsend~\cite{47BT,53Bat}. For recent development, see, e.g.
Ref.~\cite{00FONGY} and references therein. It was found that
$E(t)\propto t^{-n}$ with the decay exponent, $n$, ranging
between 1 and 5/2. This large degree of uncertainty stems from
difficulties in pinpointing the energy spectrum at scales larger
than the  energy containing scale $L$. It is also not easy to
determine how $L$ depends on time. Take for example the case of
grid turbulence in a wind tunnel. Immediately behind the grid $L$
is of the order of the mesh size. It increases however with the
distance from the grid. Downstream $L$ may saturate at the wind
tunnel diameter. In this regime the phenomenological
analysis~\cite{00FONGY} predicts $n=2$, which is a number that is
not in contradiction with experiments~\cite{00SS}. The same
prediction ($n=2$) was reached in DNS of the Navier Stokes
equation~\cite{95BO} and for the GOY shell model \cite{Lohse}.
This prediction was shown to be in agrement with numerical
simulation in which $L$ is time independent due to the special
choice of initial conditions.

For the sake of completeness we review the theoretical analysis of
\cite{Lohse}, and show that our simulations of the Sabra shell model are
in excellent agreement with its predictions.
\begin{figure}
\epsfxsize=8.5cm \epsfbox{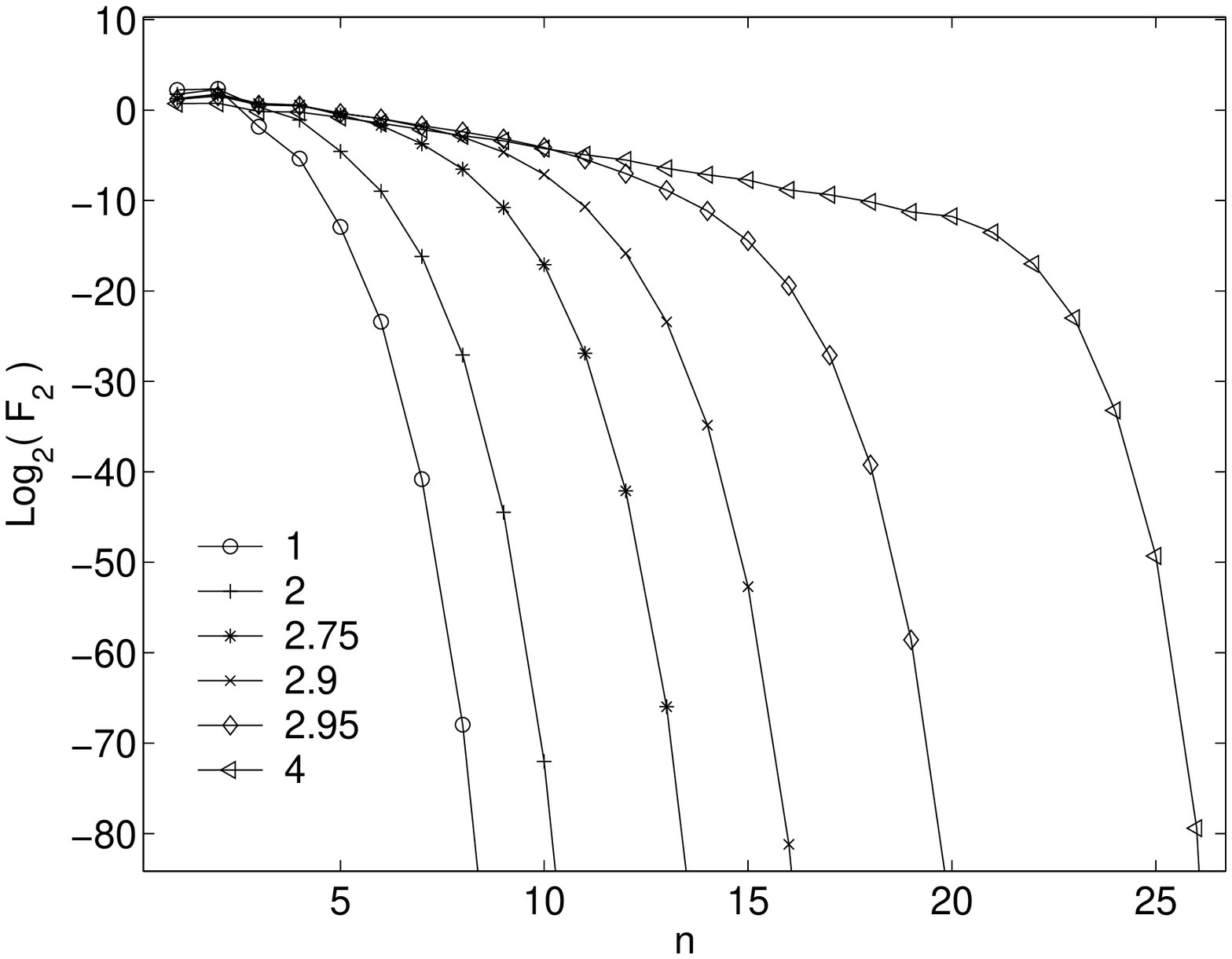} \epsfxsize=8.5cm
\epsfbox{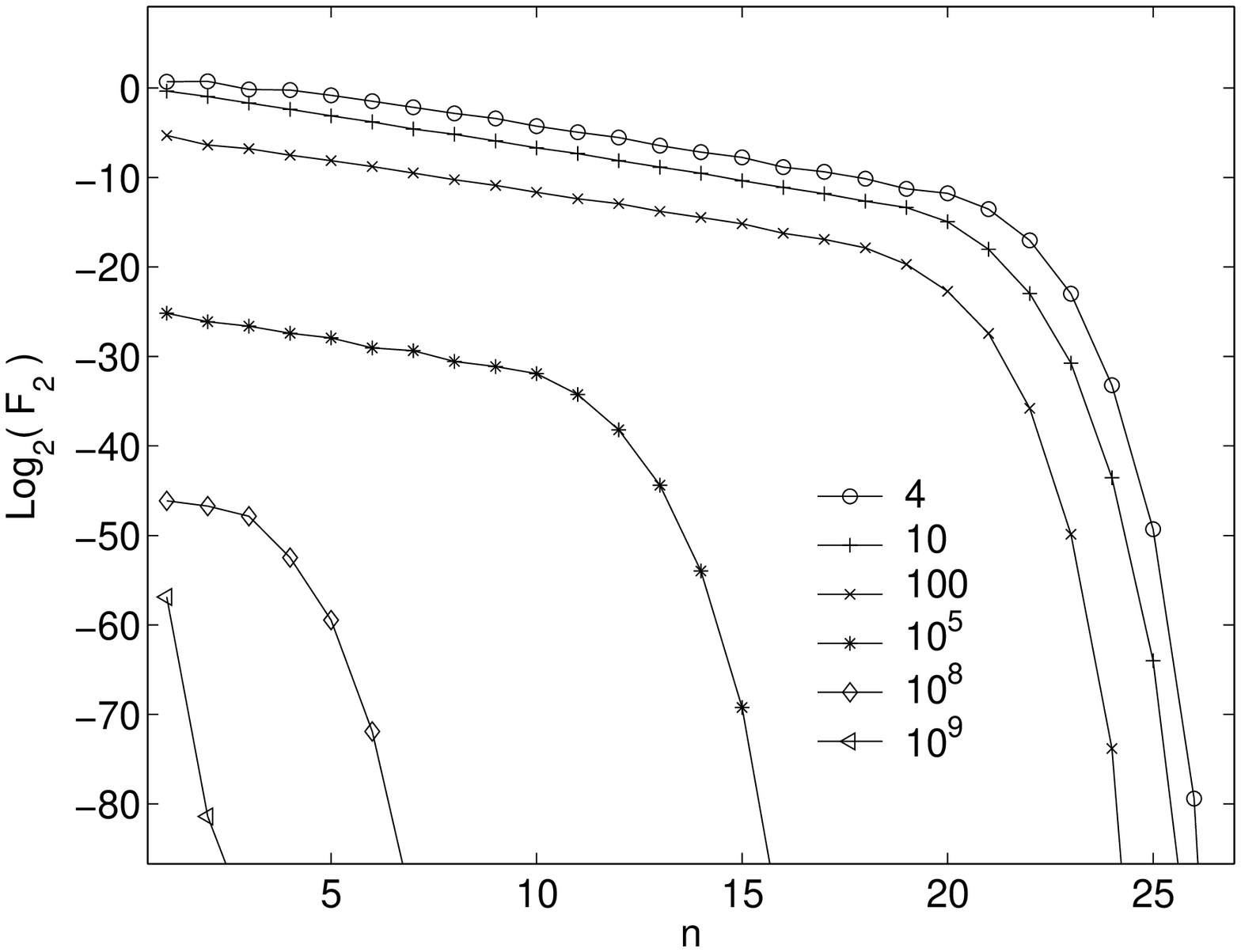} \caption{\label{f:f2-evolution}The
behavior of $F_2$. Development of the energy cascade ( upper
panel). The decay phase ( lower panel).  }
\end{figure}
Consider a decaying solution with the energy initially
concentrated, say,  in the first two shells, see
Fig.~\ref{f:f2-evolution}, upper panel. Time is measured in
natural time units $T$ which are determined by the characteristic
time of the first shell, $T = 1/5\, k_0\sqrt {E_0}$ ($E_0$ is the
total initial energy). One sees that during one $T$ the energy
cascaded down to the 6th shell, and during $2T$ down to the 12th
shell. At later times the cascade process accelerates, and the
energy goes from the 13th shell to ``infinite" shells during a
time that is roughly between $2.75\, T$ and $3\,T$. As expected,
the completion of the cascade process requires a finite time
$T_*$ of a few units $T$. In our case $T_*\approx 3 \, T$.

For $t>T_*$ the total energy of the system $E(t)$ begins to decay
with
\begin{equation}\label{eq:dE/dt}
\frac{d E(t)}{d t}= - \bar \varepsilon_{\rm i}(t)\ .
\end{equation}
Accordingly, the time dependent Reynolds number $\C R{\rm e}(t)$,
which is proportional to $\sqrt{E(t)}$, decreases, and the viscous
cutoff $k_d(t)$  is moving toward smaller shell numbers, as is
shown in Fig.~\ref{f:f2-evolution}, lower panel. For example,
$k_d(10^5T)\approx 2^{12}$, $k_d(10^8T )\approx 2^3$ and the
inertial interval almost disappeares. For larger times all the
energy is contained in the first shell, and it decays
exponentially,
\begin{equation}\label{eq:visc-decay}
E(t)\propto \exp[-2 \nu k_1^2 t]\,,
\end{equation}
following the linear part of the equation of motion for the first shell.

For intermediate times,  which in our simulations span the eight
orders of magnitude for $ 3\, T <t< 10^8 T $, the slope of plots
of log$_2 F_2(k_n)$ vs $n$ remains more or less constant. This is
a manifestations of the time independence of the scaling
exponents. Taking this as a fact, one immediately see from
Eq.~(\ref{eq:scaling-Fp}) for $p=2$ that $E(t)\propto [\bar
\varepsilon_{\rm i}(t)]^{2/3}$ and hence $\bar \varepsilon_{\rm
i}(t)\propto [E(t)]^{3/2}$. Note that this result is independent
of the precise value of the scaling exponent $\zeta_2$, anomalous
or not. Thus Eq.~(\ref{eq:dE/dt})  can be presented as
\begin{equation}\label{eq:dec-eq}
\frac{d E(t)}{d t}= - \kappa \,  [E(t)]^{3/2}\,,
\end{equation}
with a pre-factor  $\kappa$, which may be expressed via the
parameters of the shell model, $k_0$, $\zeta_2$ and $C_2$.
Approximately, $\kappa
 \simeq k_0$.  In our calculations, both Decay 1 and Decay 2,
$\kappa=0.0687$, while $k_0=0.0625$. The solution of Eq.~(\ref{eq:dec-eq}) is
\begin{equation}\label{eq:sol1}
E(t)=E_*\frac{t_*^2}{(t+t_*)^2}\,,
\end{equation}
where $t_*=2/\kappa \sqrt {E_*}$ and $E_*$ is the integration
constant.
\begin{figure}\label{Fig.2}
\epsfxsize=8.5cm \epsfbox{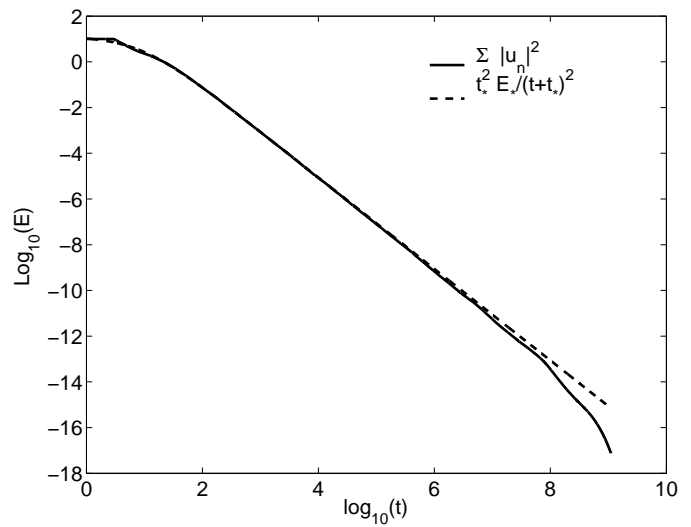} \caption{ The decay of
the total energy $E=\sum_n | u_n |^2$
 averaged over 3200 initial conditions. The dashed line corresponds to
  the decay law, Eq.~(\ref{eq:sol1}),
   with  $E_*=12.15$ and $t_*=8.35\, T$.}
\end{figure}
The results of the  numerical simulations for the total energy,
$E(t)$, (cf.  Fig.~\ref{Fig.2}, solid line), are in excellent
agreement with  Eq.~(\ref{eq:sol1}), which is shown in the figure
as a dashed line. The total energy decay was followed for 9
decades in time. After 1.5 decades of transient behavior, the
decay of total energy follows very closely the $t^{-2}$ law,
until at about 6 decades the viscous scale reaches the first
shells and the decay become exponential in agreement with
Eq.~(\ref{eq:visc-decay}).
\section{Forced structure functions}
\label{s:forced}

 In this section we present results for the forced
structure functions. As far as the scaling exponents are
concerned, there is not much novelty in this section, the
exponents are basically the same as those reported in a number of
previous publications. The new aspect stressed here is that of the
{\em coefficients} $C_p$ of the structure functions, cf.
Eq.~(\ref{eq:scaling-Sp}). We demonstrate that these coefficients
are independent of the forcing, and that the scaling form proposed
in Eq.~(\ref{eq:scaling-Sp}) is indeed universal.

To get an accurate determination of the scaling exponents and of
the coefficients, and to be able to demonstrate strong
universality, it is mandatory to fit the measured data to model
functions that contain the presumed dissipative behavior. Failing
to do so results in inaccuracies that may lead to infinite
confusions. As a first step in analysis of the data we fit the
normalized structure functions $S_p/E^{p/2}$ with the fit formula
Eq.~(44) from  \cite{98LPPPV}:
\begin{equation}\label{eq:fit_func}
P_p(k_n)=\frac{A_p}{k_n^{\zeta_p}}\Big (1+\alpha_p
\frac{k_n}{k_{{\rm d},p}} \Big )^{\mu_p}\exp
\Big[-\Big(\frac{k_n}{k_{{\rm d},p}}\Big )^x \Big]\, ,
\end{equation}
where
\[
A_p,\zeta_p\,,\ \alpha_p\,,\ \mu_p\,,\  k_{{\rm d},p}\,,\quad {\rm
or }\quad  n_{{\rm d},p}\equiv \log_2 k_{{\rm d},p}
\]
are the fit parameters and  $x=\log_{\lambda}(1+\sqrt 5)/2$ is the
exponent of the viscous range.  The parameters  $A_p,\zeta_p$
determine the behavior of $P_p(k_n)$ in the inertial interval,
$k_{{\rm d},p}$ determines the viscous cutoff. The ``auxiliary
parameters" $\alpha_p,\mu_p$ correct the behavior in the transient
inertial-viscous region.
\begin{figure}
\epsfxsize=8.5cm \epsfbox{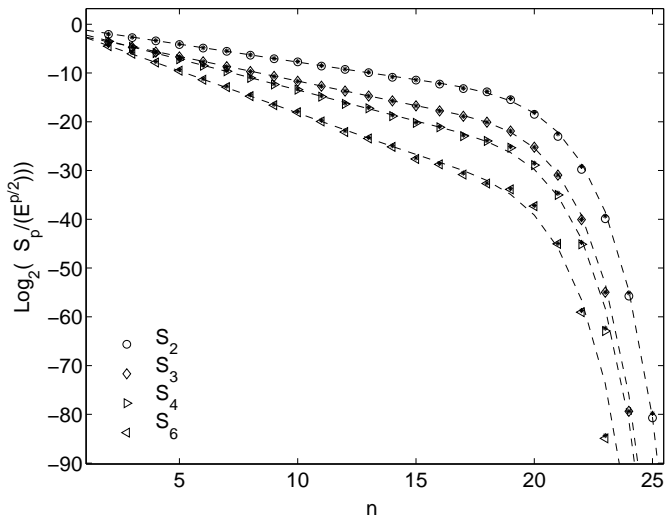} \caption{\label{f:forced}
The structure function for two types of the forcing. Large
symbols, shown in the figure  correspond to Forced 2 data. The
dashed lines are the fits for the corresponding structure
functions.  Small black dots denote Forced 1 data. The fits for
these data are not shown. Both sets were normalized by their
respective total energy.}
\end{figure}
To obtain the best fit we minimize the error function:
\begin{equation}\label{eq:fit_err}
\C E =\sqrt {\sum_n \Big (1-\frac{\log_{10} P_p(k_n)}
{\log_{10}S_p(k_n)}\Big)^2}\, ,
\end{equation}
where $S_p$ refers to the numerically obtained data. Both sets of
forced structure functions data were fit with all the shells
taken into account except the first two and the last three shells,
to minimize the boundary effects.

The quality of the fit may be seen in Fig.~\ref{f:forced}.  The
forced data are shown normalized by the respective total energy,
but not compensated by $k_n^{\zeta_p}$. Then the different
structure functions are separated and data for Force 1 and Force 2
cases may be distinguished.

 The fit procedure allows us to express the structure
function in the inertial range as $S_p=A_p E^{p/2} k_n^{\zeta_p}$.
 To calculate the coefficients $C_p$ we have now estimate the
value of the energy flux $\bar \varepsilon_{\rm f}$[ see
Eq.~(\ref{eq:scaling-Sp})]. We use the exact result for $S_3$ to
express $\bar \varepsilon_{\rm f}$ via $A_3$ and parameters of the
model Eq.~(\ref{eq:S3}). The coefficient $C_p$ of the structure
functions, other than $S_3$, may be therefore written as:
\begin{equation}\label{eq:Cp}
C_p=\frac{a_p}{[2 a_3(a-c)]^{p/3}}\ .
\end{equation}
The results are summarized in the Table \ref{tab:1}. Before we
discuss this Table, which is the central result of this paper, we
turn to the analysis of the decaying structure functions and add
their analogous results to the Table as well.

\section{Decaying structure functions}
\label{s:decay}

 In this section we present results for the
decaying structure functions, including the numerical support for
the strong universality proclaimed in Eq. (\ref{eq:scaling-Fp}).
We caution the reader (and whoever wants to repeat these
calculations in other systems, including Navier-Stokes decaying
turbulence), that the issue is fraught with sub-leading
contributions, even in the isotropic sector. One obvious
sub-leading term is provided by the rate of change of $F_p(k_n,t)$
which is coupled, via the infinite hierarchy  of equations, to
terms involving $F_{p+1}(k_m,t)$, with $m$ of the order of $n$. To
see this phenomenon clearly and to learn how to take it into
account we discuss first the case of the 3rd order structure
function which can be dealt with analytically.
\subsection{Sub-leading corrections to the scaling of decaying turbulence}
The easiest case for theoretical analysis of the scaling behavior
of the decaying structure functions $F_p(k_n,t)$ is the case
$p=3$, for which in the forced case $S_3(k_n)$ is known
exactly~\cite{98LPPPV}. For  simplicity we will discuss here the
helicity-free case, for which
\begin{equation}\label{eq:S3}
S_3(k_n)= \bar \varepsilon _{\rm f}/k_n (c-a)\ .
\end{equation}
 We will show now that in the decaying case strong universality is
 realized, but only well within the inertial range. In the vicinity of the
 energy containing scales there are significant sub-leading corrections
 caused by the time dependence of $\bar\varepsilon_{\rm i}(t)$.

To find these corrections, consider the equation of motion of the
2'nd order structure in the inertial interval (i.e. for $k_n\ll
k_d$) \{Eq.~(9) of Ref.~\cite{98LPPPV}\}:
\begin{equation}\label{eq:budget}
\frac{d F_2(k_n,t)}{2 k_n d t}=a\lambda F_3(k_{n+1},t)+
bF_3(k_{n},t)+ \frac c\lambda F_3(k_{n-1},t)\ .
\end{equation}
In the stationary case, the LHS of this equation vanishes and
Eq.~(\ref{eq:S3}) is a solution. In the decaying case, let
\begin{eqnarray}\label{eq:corr}
F_3(k_n)&=&  F_3^{(0)}(k_n,t)+  \delta F_3(k_n,t)\,, \\
\nonumber F_3^{(0)}(k_n)&=& \frac{\bar \varepsilon _{\rm
i}(t)}{k_n (c-a)} \ .
\end{eqnarray}
In the intermediate time regime $F_2(k_n,t)\propto (t+t_*)^{-2}$,
\begin{equation}
d F_2(k_n,t)/ 2 k_n d t \approx
   - F_2(k_n,t)/k_n\,(t+t_*)
\ . \label{eq:correl}
\end{equation}
Comparing with Eq. (\ref{eq:budget}), we see that the leading solution for $F_3$\
gains a sub-leading term
 $\delta F_3(k_n,t)\propto k_n^{-(1+\zeta_2)}$ or more precisely
\begin{equation}
\delta F_3(k_n,t) \approx  - \frac{ F_2(k_n,t)}{(t+t_*)\,
k_n(a\, \lambda ^{-\zeta_2}+ b + c \lambda ^{\zeta_2})}\ .
\end{equation}
Notice, that $  F_3^{(0)}(k_n,t)$ and $\delta F_3(k_n,t)$ have the
same time dependence, $\propto (t+t_*)^{-3}$, but different
scaling. As a result, their ratio is time independent; the sub-leading
term does not become relatively smaller in time. On the other hand it
decays relatively to the leading term as $k_n$ increases:
\begin{equation}\label{eq:corr2}
\frac{\delta F_3(k_n,t)}{ F_3^{(0)}(k_n,t)}\propto
\frac{1}{\lambda^{n \zeta_2}}\ .
\end{equation}

Although the 3rd order structure function is easiest to handle,
it is clear that there will always be a subleading term added to
$F_p$ from the time derivative of $F_{p-1}$ which appears in the
infinite hierarchy of equations. Since these equations always
have $k_n$ on the RHS, one can immediately guess the general form
of the correction to scaling for $F_p$, i.e.
\begin{equation}\label{eq:correl2}
\frac{\delta F_p (k_n,t)} {F_p (k_n,t)} \propto \frac 1 {\lambda
^{n(1+\zeta_{p-1}- \zeta_p)}}\ .
\end{equation}
This means that the sub-leading term of the $p$-order structure
functions decreases toward small scales roughly as $\lambda
^{-2n/3}$ for ``normal" K41 scaling, and somewhat slower for
anomalous scaling. Strong universality of the turbulent statistics
is thus expected only deeply in the inertial interval.

\subsection{Numerical results}

\begin{figure} \epsfxsize=8.5cm \epsfbox{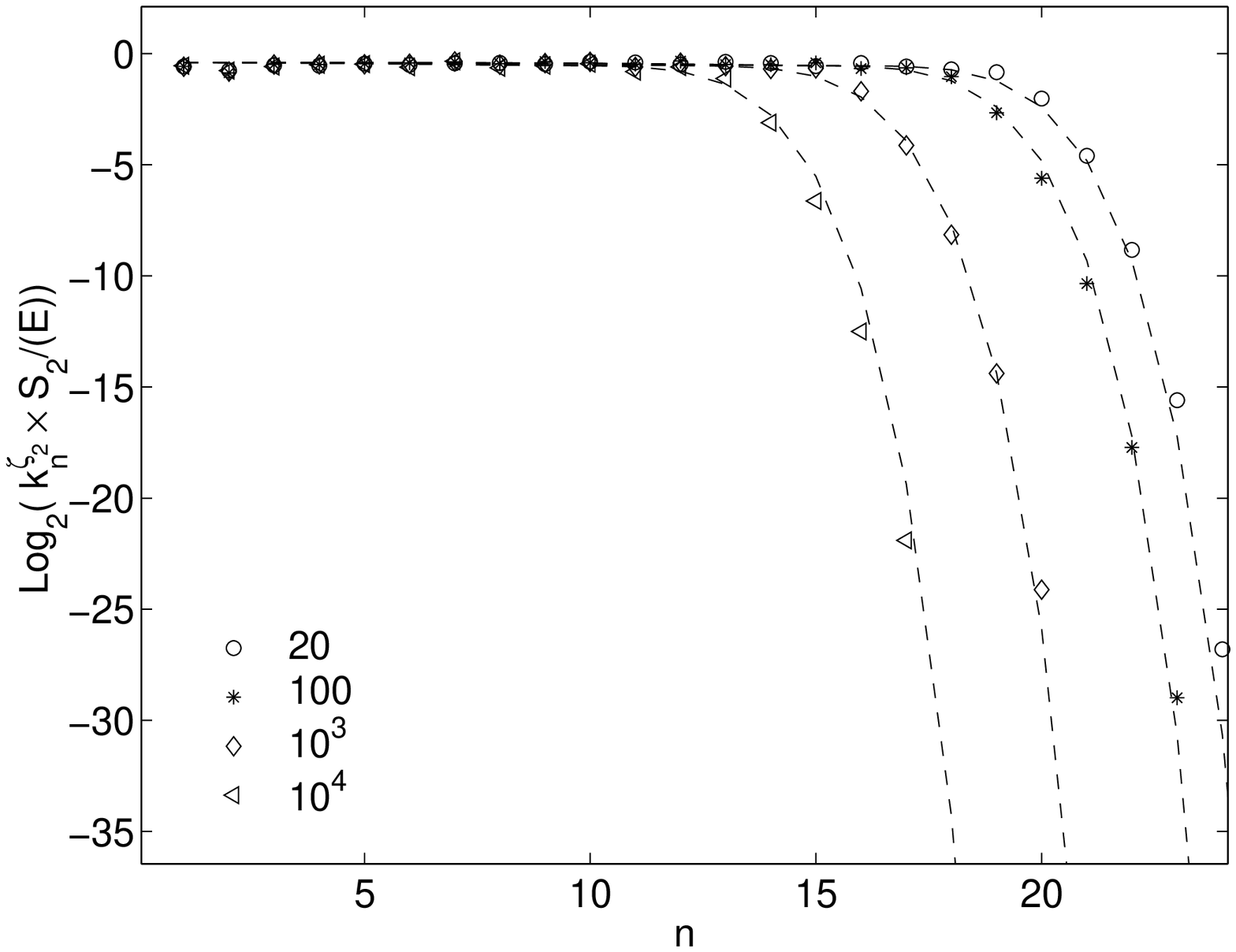}
 ~~~~~~\epsfxsize=8.5cm \epsfbox{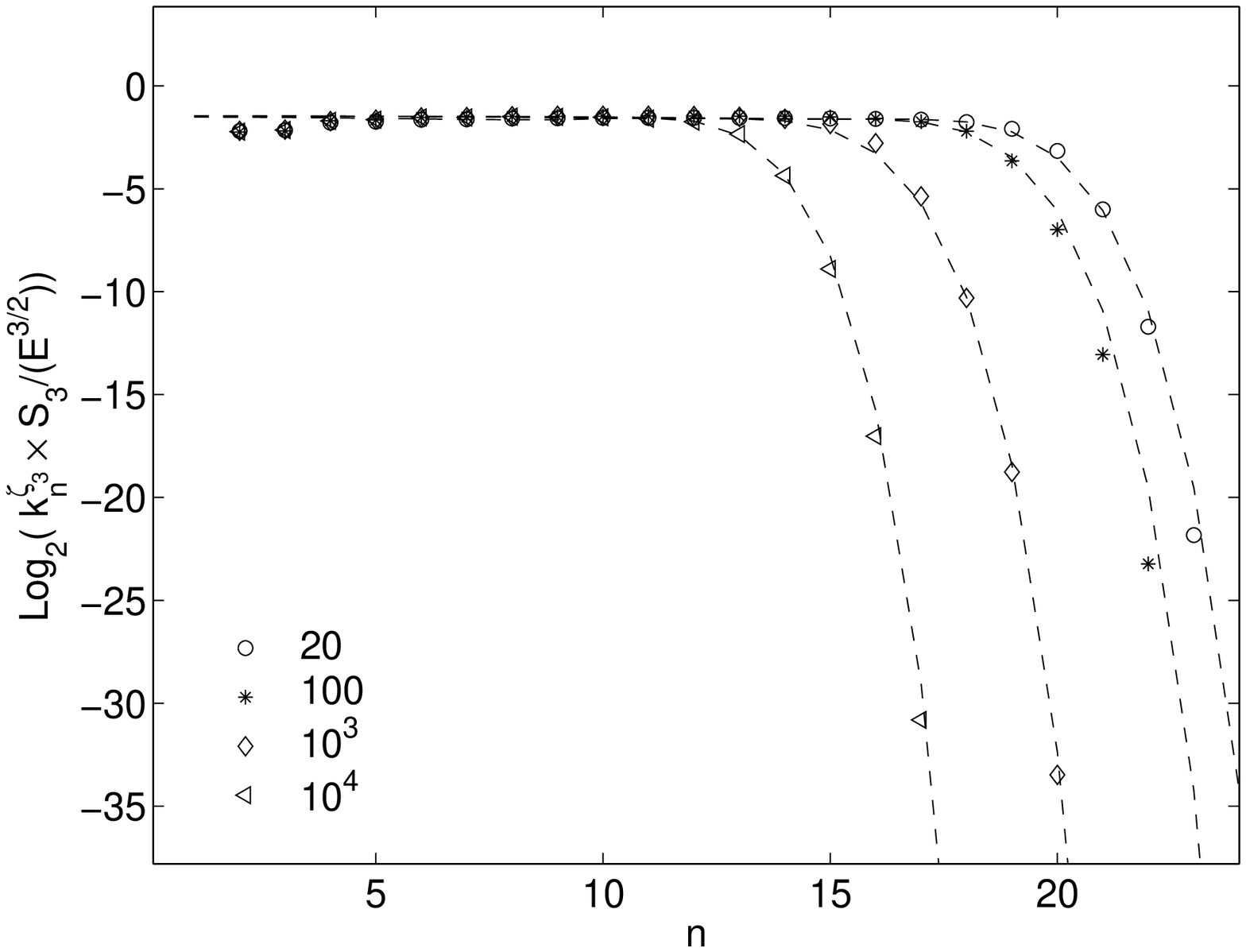}
 \epsfxsize=8.5cm \epsfbox{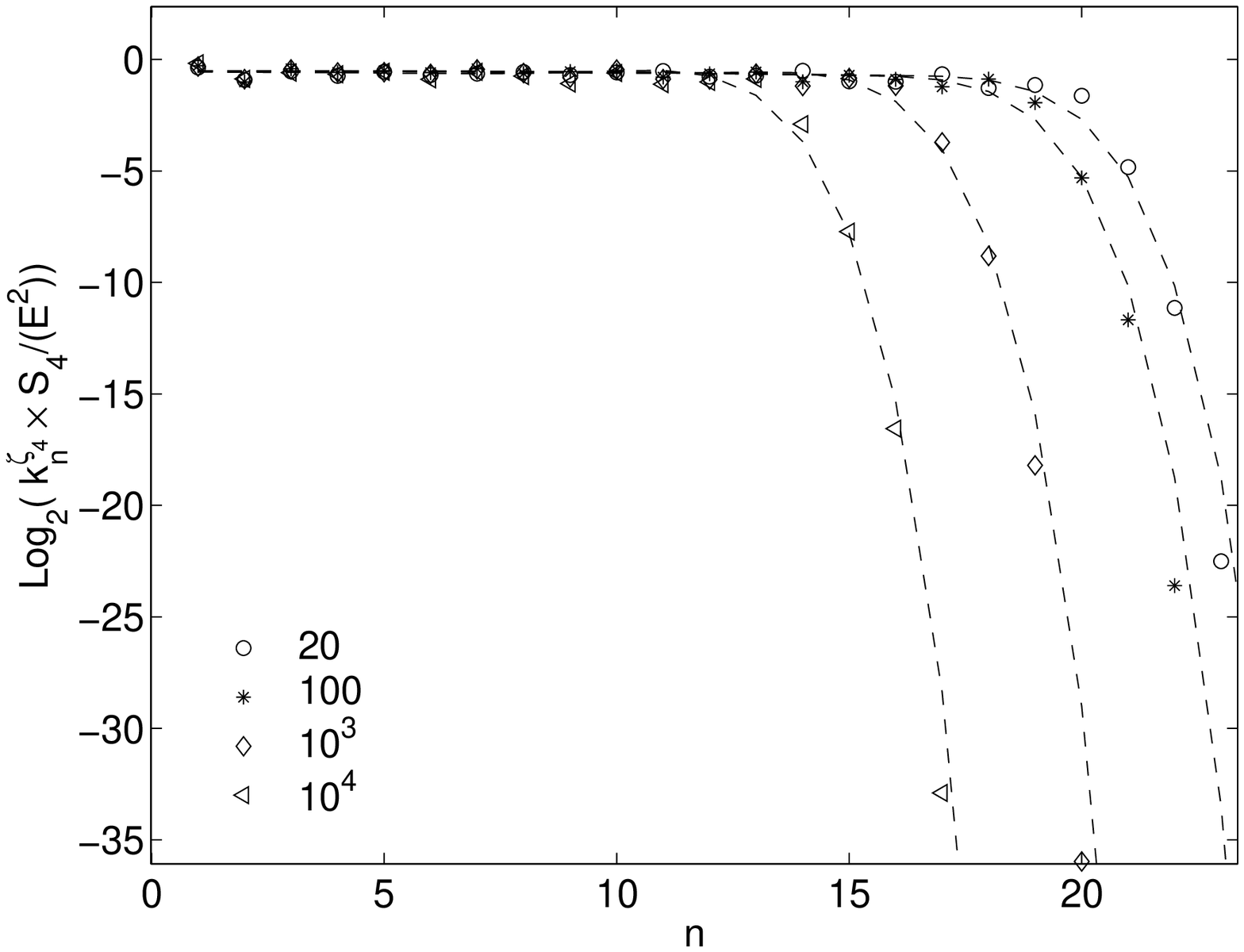} ~~~~~~
\epsfxsize=8.5cm \epsfbox{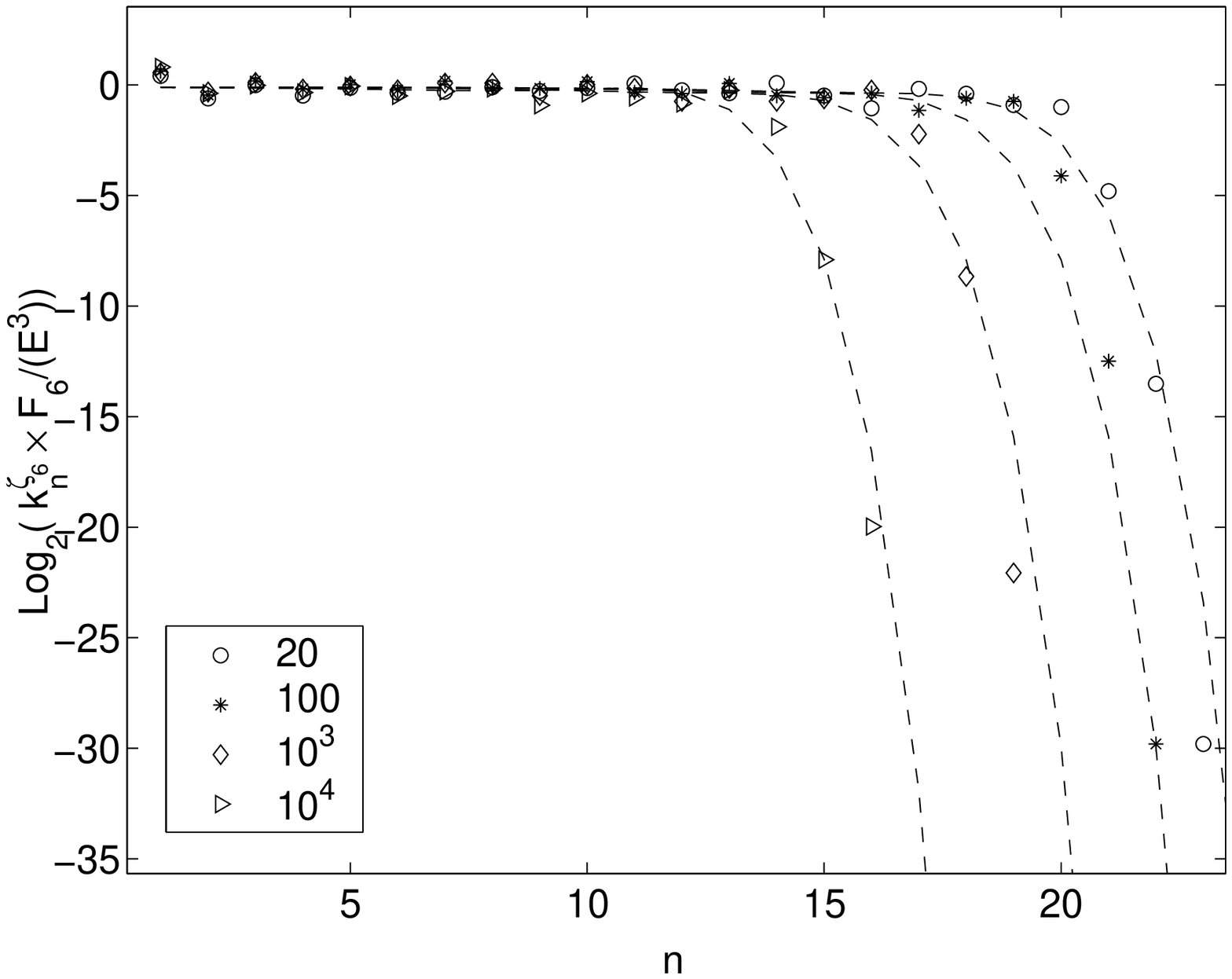} \caption{\label{f:init} The
normalized compensated decaying structure functions for Decay 1
data averaged over 13200 initial conditions  and over one tenth of
the decade in time. The symbols show the calculated data at
different times( defined in the legend). The dashed lines denote
the fits for the corresponding structure functions.
 The best fit scaling exponents, used for
  the compensation ($\zeta_2=0.728,
\zeta_3=1., \zeta_4=1.254$ and $\zeta_6=1.72$)  are the same as in
the forced case. The Decay 2 data show similar behavior.}
\end{figure}

 All calculated statistical objects were normalized
by the total energy,  $F_p(k_n,t)/E^{p/2}(t) $.  All the
normalized, compensated decaying structure functions show a
plateau. To enrich the statistics the data were first normalized
by the total energy and then averaged over one tenth of a temporal
decade. For the same reasons that were explained in the case of
the forced objects, the fit region for the decaying structure
functions was chosen from $n=3$ to $n = n_{ {\rm d},p}+5$ for the
Decay 1 data and from $n=5$ to $n = n_{ {\rm d},p}+5$ for the
Decay 2. For $t=10^5 \, T $ $n_d\approx 7 $ and only very few
shells may be considered as the ``inertial interval". Therefore
the fit parameters become unreliable. The quality of the fit  can
be seen in Fig. \ref{f:init} for $t$ between $20\, T $ and $10^4\,
T$. For $t \le 20\,T $ the flux equilibrium cannot be guaranteed
and the coefficients $C_p$ and the exponents may be not universal.
This is definitely the case for $t  \le 10\, T$. The decaying
structure functions are plotted normalized by the total energy and
compensated to emphasis the fact that main effect is the shift of
$k_{{\rm d},p}$.
\Onecol

\begin{table}\caption{\label{tab:1}
 Universal coefficients $C_p$ and scaling exponents measured from
 "the best fit" on numerical data.
 The auxiliary fit parameters in Eq. (\ref{eq:fit_func}) are
 found to be in the intervals  $\alpha_p\sim 0.5 \div 2$
 and $\mu_p\sim 0.6 \div 2.3$. The error  bars for each parameter
 correspond to the error function $\C E$, Eq.~(\ref{eq:fit_err} )
 equal
 $\simeq\sqrt 2\C E_{\rm min}$ with all other parameters
  set to their optimal values.}
\begin{tabular}{||l|c||c|c|c||c|c|c||c|c|c||}
\hline\hline
 &Time & $C_2$ &$\zeta_2$ & $n_{\rm d,2}$& $C_4$
 &$\zeta_4$ & $n_{\rm d,4}$ & $C_6$&$\zeta_6$
 & $n_{\rm d,6}$ \\
\hline\hline Forced 1& &~$0.73 \pm 0.07$~&~$ 0.728\pm 0.006 $ &~
17.0 ~& ~$0.60  \pm 0.08$~& ~ $1.254\pm 0.008$ ~& ~16.24~
&~ $0.62 \pm 0.15$~ &~$ 1.72\pm 0.01 $ ~& ~ 15.87  \\
\hline Forced 2 & & $0.73 \pm 0.07$&$ 0.728\pm 0.006 $ & 17.0 &
$0.60 \pm 0.08$&  $1.254\pm 0.008$ & 16.24
& $0.61  \pm 0.15$ &$ 1.72\pm 0.01 $ &  15.87  \\
\hline \hline

Decay 1 & 20 &$0.72 \pm 0.05 $&$ 0.728\pm 0.006 $ & 17.2 & $0.62
\pm 0.06$& $1.254\pm 0.008$ & 17.0
& $0.79  \pm 0.15$ &$ 1.72\pm 0.01 $ &  16.4   \\
\hline  & 100 &$0.72 \pm 0.06 $&$ 0.728\pm 0.006 $ & 16.2 & $0.62
\pm 0.07$& $1.254\pm 0.008$ & 16.0
& $0.79  \pm 0.15$ &$ 1.72\pm 0.01 $ &  15.0   \\
\hline  & $10^3$ &$0.72 \pm 0.07 $&$ 0.728\pm 0.006 $ & 13.5 &
$0.61 \pm 0.08$& $1.255\pm 0.008$ & 13.2 & $0.77  \pm 0.15$ &$
1.72\pm 0.01 $ &  13.0   \\
\hline & $10^4$ &$0.72 \pm 0.08 $&$ 0.728\pm 0.006 $ & 11.0 &
$0.61 \pm 0.09$& $1.256\pm
0.008$ & 10.2 & $0.75  \pm 0.15$ &$ 1.73\pm 0.01 $ &  9.8   \\
\hline 
Decay 2  &  100 &$0.73 \pm 0.1 $&$ 0.73\pm 0.01 $ & 15.5 & $0.6
\pm 0.2 $& $1.25\pm 0.01$ & 15.2
& $0.70  \pm 0.20$ &$ 1.72\pm 0.02 $ &  14.3    \\ \hline 
 & $10^3$ &$0.74 \pm 0.1 $&$ 0.73\pm 0.01 $ & 13.3 &
$0.6 \pm 0.15 $& $1.26 \pm 0.01$ & 13.1 & $0.66  \pm 0.20$ &$ 1.72
\pm 0.02 $ &  12.4
 \\ \hline  
& $10^4$ &$0.72 \pm 0.1  $&$ 0.73\pm 0.01 $ & 10.7 & $0.6 \pm 0.2
$& $1.25\pm
0.01 $ & 10.5 & $0.72  \pm 0.25$ &$ 1.73\pm 0.03 $ &  9.8  \\
 \hline\hline
\end{tabular}
\end{table}
\Twocol
\section{Summary and discussion}
\label{s:sum}

In summary, we analyzed, using the Sabra shell model of
turbulence, ``forced" structure functions $S_p(k_n)$ for two
types of forcing, and ``decaying" structure functions
$F_p(k_n,t)$ for two types of initial conditions, (called Decay~1
and Decay~2). For Decay 1 we considered  four times, which differ
in order of magnitudes ($t=20\, T , \, 10^2\, T,\, 10^3\, T,$ and
$ 10^4\, T$, where $T$ is the characteristic time of the 1st
shell). For Decay~2 we considered three different times $t=10^2\,
T,\, 10^3\, T,$ and $ 10^4\, T$. In all these cases we found the
scaling exponents, $\zeta_p$, and the dimensionless amplitudes,
$C_p$, for the three even orders $p=2,\, 4$ and $6$. The results
are collected in Table 1 together with our estimates of the error
bars.

We can state that our results {\em support the conjecture of
strong universality} within the numerical accuracy. Concerning
the fact that before the data analysis presented above the raw
results contained objects differing by orders of magnitude, the
degree of precision of the identity of the amplitudes $C_p$, and
the exponents $\zeta_p$, shown in Table 1 should be taken very
seriously. We propose that all the results presented by previous
authors with negative indications about universality (even of the
weak type) stem from problems in handling the corrections to
scaling, either from anisotropy or from dissipative or other
boundary effects.

It should be stressed at this point that the strong universality
observed here is not expected in the much simpler problem of
turbulent advection. The difference stems from the linearity of
the advection problem vs. the nonlinearity of the Navier-Stokes
problem and its shell counterpart. In the linear advection problem
one finds equations for the statistical objects that decouple for
each order $p$. The independent $p$-order equations determine the
anomalous scaling exponent $\zeta_p$ from solvability conditions,
leaving the amplitude $C_p$ to be found by matching the scale
invariant correlation function in the inertial interval with its
non-universal ``boundary conditions" at energy contained scales.
The amplitudes $C_p$ depend therefore on the details of the
non-universal forcing, and the statistics of the turbulent
advection problem may exhibit weak universality only. In contrast,
the nonlinearity of the Navier Stokes equations and their
shell-model counterparts leads to coupled, hierarchical equations
for all the $p$-order statistical objects that have to be solved
for simultaneously.  This rigid structure allows much less freedom
than the linear advection problem, leading to the possibility of
strong universality.

Finally, we comment on a possible theoretical support for the
strong universality conjecture. We propose that a necessary
condition for  strong universality is the locality of interaction,
which allows to formulate (see Refs.~\cite{98LP,98BLP,98BLPP}) the
hierarchy of equations in terms of inertial-range objects only.
The locality of energy transfer over scales, which is built in the
shell models of turbulence, is an assumption in the
Richardson-Kolmogorov cascade picture of turbulence, see,
e.g.~\cite{75MY,41Ka,95Fri}. The locality of interaction was
demonstrated in Ref.~\cite{95LP}, using the Belinicher-L'vov
transformation of the Navier-Stokes equations~\cite{87BL}, which
allows to eliminate from the theory the sweeping effect. Once we
have a theory in terms of inertial-range objects, it is quite
acceptable that amplitudes should be universal as well, up to an
overall single parameter which is the energy flux per unit time
and mass. An elaboration of these ideas will be presented in a
future publication. At this point we finish with the conjecture
that strong universality is a property shared also by
Navier-Stokes turbulence.

\begin{acknowledgments}
This research was supported in part by Israel Science Foundation
administered by the Israeli Academy of Science and the European
Commission under the TMR network ``Nonideal Turbulence".

\end{acknowledgments}

\end{document}